\documentstyle[12pt,aasms4]{article}
\input epsf
\begin{document}

\title{A search for correlations of TeV $\gamma$-rays with \\
ultra-high energy cosmic rays}


\author{C. W. Akerlof\altaffilmark{3},
S. Biller\altaffilmark{4},
P. Boyle\altaffilmark{2},
J. Buckley\altaffilmark{1},
D. A. Carter-Lewis\altaffilmark{5},
M. Catanese\altaffilmark{5},
M. F. Cawley\altaffilmark{6},
V. Connaughton\altaffilmark{1,10},
D. J. Fegan\altaffilmark{2},
J. Finley\altaffilmark{7},
J. Gaidos\altaffilmark{7},
A. M. Hillas\altaffilmark{4},
F. Krennrich,\altaffilmark{5},
R. C. Lamb\altaffilmark{5},
R. Lessard\altaffilmark{2},
J. McEnery\altaffilmark{2},
G. Mohanty\altaffilmark{5},
N. A. Porter\altaffilmark{2},
J. Quinn\altaffilmark{2},
A. Rodgers\altaffilmark{4},
H. J. Rose\altaffilmark{4},
F. Samuelson\altaffilmark{5},
M. S. Schubnell\altaffilmark{3},
G. Sembroski\altaffilmark{7},
R. Srinivasan\altaffilmark{7},
T. C. Weekes\altaffilmark{1}, and
J. Zweerink\altaffilmark{5}.}

\altaffiltext{1}{CfA - Whipple Observatory, P.O. Box 97, Amado,  AZ 85645}
\altaffiltext{2}{University College Dublin, Physics Department, Belfield, 
Dublin 4, Ireland}
\altaffiltext{3}{University of Michigan, Department of Physics, Randall 
Laboratory, Ann Arbor, MI 48109-1120}
\altaffiltext{4}{University of Leeds, Department of Physics, Leeds LS2 
9JT, United Kingdom}
\altaffiltext{5}{Iowa State University, Department of Physics \& 
Astronomy, Osborn Drive, Ames, Iowa 50011}
\altaffiltext{6}{St.Patrick's College, Department of Experimental Physics,
Maynooth, Co Kildare, Ireland}
\altaffiltext{7}{Purdue University, Department of Physics, Lafayette, IN 47907}
\altaffiltext{10}{Contact address: NASA Marshall Space Flight
Center, ES 84, AL 35812}



\begin{abstract}
A search was conducted for TeV $\gamma$-rays emitted from the direction of the
ultra-high energy cosmic ray detected by the Fly's Eye Experiment with $E\sim 3
\times 10^{20}$ eV.  No enhancement was found at a level of $10^{-10}
\gamma$/cm$^2$-sec for $E>350$ GeV.  This upper limit is consistent with 
theoretical estimates
based on topological defects as sources of UHE cosmic rays. An upper limit was
also set for the flux of TeV gamma rays from 3C147, the most prominent AGN in
the error box. 
\end{abstract}
\vspace{.5in}

\section{Introduction}
The surprising discovery of ultra-high energy (UHE) cosmic rays 
with $E>10^{20}$
eV poses significant questions about how such particles can reach energies
substantially in excess of the Greisen-Zatsepin-Kuz'min cutoff (Greisen
1966; Zatsepin \& Kuz'min 1966) imposed by interactions with the cosmic
microwave background radiation.  If these particles are accelerated in
relativistic shock fronts in a manner similar to the standard models for
lower energy cosmic rays (Blandford 1978; Legage 1983; Bell 1978), the 
physical
constraints are difficult to reconcile with what we know about possible
acceleration sites on distance scales of 40 Mparsecs.  This has suggested 
to several
authors (Bhattacharjee 1990, Bhattacharjee {\em et al.} 1992, Aharonian {\em
et al.} 1992, Sigl {\em et al.} 1994) that a new ``non-acceleration"
mechanism is at work such as the decay of GUT scale topological defects with
characteristic masses of the order of $10^{25}$ eV.  A second possible
origin for UHE cosmic rays is gamma-ray bursts (Milgrom \& Usov 1995, Vietri
1995, 1996, Miralda-Escud\'{e} \& Waxman 1996); Waxman (Waxman 1995) has
pointed out that UHE cosmic rays and gamma-rays from GRBs have comparable
fluences at the Earth, the possible consequence of energy equipartition. 
Although no one knows what the physical mechanism of gamma-ray bursts really
is, it remains conceivable that these objects could also accelerate charged
particles to ultrarelativistic energies.

Since the lifetime of superenergetic cosmic rays is limited to $10^8$ years
and the magnetic rigidity is extremely high, it is likely that the sources
of this radiation lie close to the arrival directions measured on Earth. 
This suggests that a search for TeV gamma-ray counterparts might offer
some chance of detection.  The optical depth for TeV photons (Nikishov 1962; 
Gould 1966; Stecker 1992; Biller 1995) is considerably
greater than the range of UHE cosmic rays so attenuation is negligible.
Thus, such gamma radiation should be an excellent probe of higher
energy phenomena which would be otherwise be opaque.  The starting point for 
guessing the gamma-ray flux is
the cosmic ray spectrum measured by the Fly's Eye Experiment (Bird {\em et
al.} 1994): $$J(E) = 5.13 \times 10^{21} ~~ (E/1~{\rm ev})^{-3.07}~~cm^{-2} 
s^{-1} sr^{-1} eV^{-1}~~ (E>10^{17} {\rm eV}).$$ 

\noindent
For energies greater than $3 \times 10^{20}$ eV, the extrapolated integral
flux is $1 \times 10^{-21}$ particles/cm$^2$-s-sr.  For the one event
measured at the end of the spectrum, the corresponding flux from a point
source is $6 \times 10^{-20}$ particles/cm$^2$-sec.  A plausible assumption,
based on the behavior of AGNs, is a constant $\nu F_\nu$ distribution
extending downwards in energy to the TeV range.  If the total available 
energy is partitioned roughly equally between gamma rays and cosmic rays, the
anticipated gamma-ray flux would be in the neighborhood of $6 \times
10^{-11}$ $\gamma$/cm$^2$-sec at $3 \times 10^{11}$ eV.  This value is in
the range of sensitivities achievable with the Whipple gamma-ray telescope 
at Mt.~Hopkins, AZ (Reynolds {\it et al.} 1993) which can detect the flux 
from the Crab Nebula ($\approx 10^{-10} \gamma$/cm$^2$-sec) with a significance 
of 7 $\sigma$ in one hour.

These considerations led us to conduct an exploratory experiment to see if
an enhancement of gamma rays could be detected from the direction of the
Fly's Eye UHE cosmic-ray event (Bird {\em et al.} 1995).  This particular
event was selected because the error box was small and the position on the
sky was convenient for observations at small zenith angle where the 
atmospheric \v{C}erenkov technique is most sensitive.  The celestial
coordinates of this event were $$\alpha (1950) = 85.2^\circ \pm
0.5^\circ,~~\delta (1950) = 48.0^\circ \pm 6.0^\circ$$ 

\section{Observations}
The object of this experiment was to locate a possible point source of TeV
radiation correlated with the direction of the Fly's Eye event. Normally,
TeV observations at Whipple are conducted with accurate {\em a priori}
knowledge of source locations.  However, it is possible, using techniques
akin to computer tomography, to reconstruct an unknown point source location
from statistical analysis of the data as was first shown in a paper by
Akerlof {\em et al.} (1991).  This technique has been
refined further and used to search for TeV photons from gamma-ray bursts
(Connaughton {\em et al.} 1997a) and supernova remnants (Buckley {\em et al.} 1997). 
 
\begin{figure}
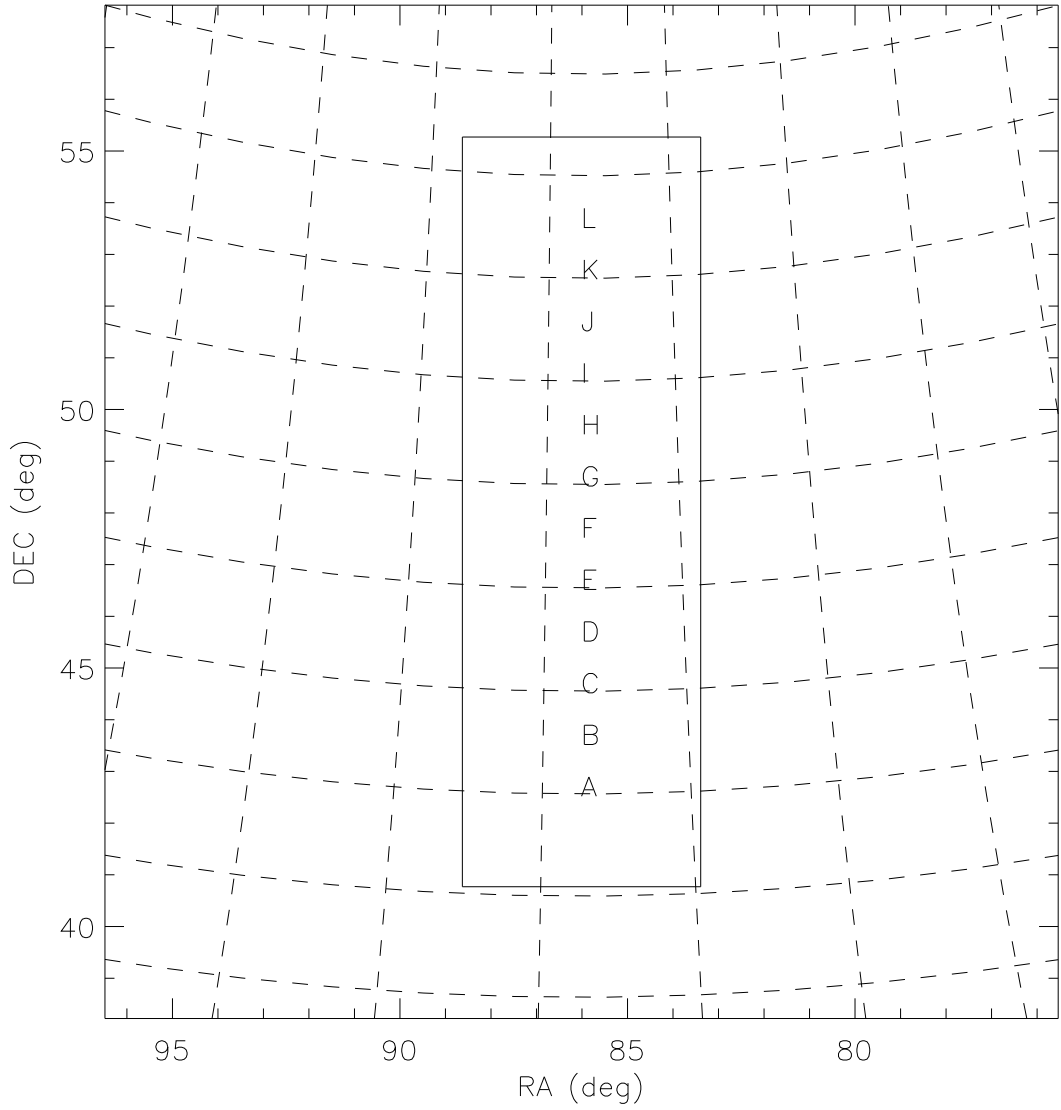

\caption{Whipple coverage of the area surrounding the Fly's Eye event.  Each 
position is centered on RA=$86.01^\circ$ and separated from
the next in declination by $1^\circ$.  Since the field-of-view is
$3.5^\circ$, there is some overlap between adjacent positions.
\label{fig:coverage}}
\end{figure}

Because of the uncertainty in the location of the topological defect, 12
overlapping regions, each centered on the Right Ascension $86.01^\circ$
(J2000), were observed with the 10m reflector. Figure~\ref{fig:coverage}
shows the region of sky covered by the Whipple observations.  The letters in
the box indicate the central point of each region.  The declinations range
from $42.51^\circ$ (Position A) to $53.51^\circ$ (Position L) in 1$^\circ$
increments so that, with the $3.5^\circ$ field-of-view of the camera, some
overlap occurred between adjacent regions. Each position was observed for
two 28 minute periods.  Observations were made during 4 nights in December
1995.  The data rate of an air shower \v{C}erenkov telescope is affected by
telescope elevation and sky conditions and hence the sensitivity and energy 
threshold of the survey varied with position. 

\section{Analysis}
Generally, observations of a point source whose location is known are analyzed
by searching for excess gamma-ray candidates from the source direction
compared to a nearby patch of sky.  Control observations of a position offset
in right ascension by 28 minutes from the source location are made with the
telescope at the same elevation as the source observations, and the excess of
selected events from the source observations (ON data) relative to the control
(or OFF source) data  gives a measure of the photon flux from the source. In
this analysis, the coordinates of the source are unknown and there are no
control observations. One must, therefore,  assume that each location on the
sky is a potential source, and look for an unexpectedly large number of
gamma-ray-like events in each of the 28 minute scans from some point in the
field-of-view. 

Standard routines (Reynolds, {\em et al.} 1993) were used to flat-field
and parameterize the data. Events in the data files are comprised of the
digitized signals registered by the 109 photomultiplier tubes in the focus box
of the 10m reflector.  For each shower produced, a moment-fitting analysis 
is used to obtain a set of image parameters
characterized by {\em width,~length,~light~concentration,} and {\em size} 
(total number of digital counts).  These parameters represent the two 
angular aspects of the shower light distribution, its compactness and total 
energy,
respectively. A combination of these image parameters has proven effective 
in discriminating against the hadronic
background by selecting only those events with the appropriate direction for 
a particular source and the shape characteristic of gamma-ray showers.  
The Supercuts technique described in (Reynolds, {\em et al.} 1993) rejects
99.7\% of the recorded background while keeping 50\% of the gamma rays. In
this analysis, gamma-ray-like events are selected on the basis of image
shape, using {\em width} (semi-minor axis of ellipse) and {\em length} 
(semi-major axis) cuts.  The development of image selection criteria and 
assessment of
non-source-centered capabilities of the 10m reflector are given in 
(Connaughton {\em et al.} 1997b). In this analysis gamma-ray-like events are 
selected suing the following Supercuts shape criteria:

\begin{eqnarray} 
0.073^\circ &  <   width  &  <   0.15^\circ \nonumber  \\
0.16^\circ  &   <   length &  <   0.30^\circ \nonumber
\end{eqnarray}

\noindent
where the width and length are the semi-minor and semi-major axes of the 
elliptical image fitted to each event.  In addition, a minimum size of 400 
dc (approximately 400 photo-electrons) is required, corresponding to an 
energy threshold of around 350 GeV.

The orientation of the ellipse fitted to each image is represented by its
major axis, and the most likely point-of-origin of the shower progenitor on
the field-of-view lies on this axis at a distance $d$ in degrees related to
the ellipticity of the image: 

\begin{equation}\label{eq:poo2}
d = 1.7 - 1.7 (width/length)
\end{equation}

This algorithm yields two points, one on either side of the center of the
image, and is considered to be accurate to about $0.3^\circ$ either side of
each point (Akerlof {\em et al.} 1991, Connaughton {\em et al.} 1997a). A 
grid of bins $0.1^\circ \times 0.1^\circ$
in size is constructed to cover the field-of-view of the camera and beyond. 
The grid extends $3^\circ$ each side of the center so that the sensitivity
of the technique outside the geometrical field-of-view can be exploited.  
Each bin which lies within $0.3^\circ$ of either point-of-origin for an
event is incremented. 

The two data files taken on each position  comprise the `ON' source data. 
Control, or `OFF source', data are obtained by averaging the grid bin
occupancies of `ON' files taken at similar elevations.  Three groups of
control data were defined: 5 of the 24 observations at telescope elevations 
below $64^\circ$, 12 between $64^\circ$ and $71^\circ$, and the remaining 7 at
higher elevations. The excess at any grid point $(i,j)$ in the ON data is
found relative to the corresponding point in the control observations using
the equation: 

\begin{equation}
\sigma_{i,j} = 
{ (N_{ON} - p \times N_{OFF}) \over 
\sqrt{(N_{ON} + p \times N_{OFF}/N_{BG})} }
\end{equation}

\noindent 
where $N_{BG}$ is the number of observations that were averaged to make up 
the background contour map.  A
normalizing factor {\em p} is applied to account for the differences in the 
durations of the ON and OFF observations.  In figure 2, the resulting 
contours on the grid
represent the significances of the excess of photon-like events over 56
minutes from each of the positions observed. 

\section{Results}
Figure~\ref{fig:burst3} shows the contour plots for positions D to I, 
typical of all 12 observations.  The
contours begin at $1 \sigma$ and increment in $1 \sigma$ steps.  Given that
there are no significant excesses in any of the bins in the ON data relative
to the control data, one can calculate an upper limit to the flux from each
of the positions observed.  The collection area above 350 GeV of the 10 m
telescope for a source in the center of the camera is $5.4 \pm 0.9 \times
10^8$ cm$^2$ (Connaughton, {\em et al.} 1997b).  This is larger than the 
collection area given in,
for example, (Reynolds {\em et al.} 1993), because of the less restrictive
orientation criteria applied to these non-source-centered observations.
Using the total number of shape-selected events in the ON and control files,
the $99.9\%$ maximum likelihood value for emission from all positions are
presented in Table \ref{tab:results}.  The errors reflect the uncertainty in
the collection area of the 10m reflector and the statistical nature of
variation in selected event rates within each control data group. These
limits apply to  emission above 350 GeV from a source in the center of the
camera. In calculating the limits from any other point in the field-of-view,
a scaling factor must be used to account for the decreasing gamma-ray
efficiencies away from the center of the camera (Connaughton, {\em et al.} 
1997b).  A lower flux
upper limit is derived for sources which might lie at the camera's center
than for those at the edge of the camera. The upper limits as a function of
source offset are shown in Figure \ref{fig:tdeful}.  The two outer lines
represent the lowest and highest limits that can be set, the middle line
shows the limits with the smallest error bars (the most homogeneous control
group), the difference being due to the statistical variations in event
rates and the diminishing efficiency of Supercuts with decreasing telescope
elevation. 

\section{Interpretation}
The lack of a detectable gamma-ray signal from the direction of the Fly's Eye
ultra-high energy cosmic ray event does not lead to any clear-cut conclusion.
First of all, the sky coverage was limited to the cosmic ray error box alone
which does not include effects of possible curvature of the trajectory by
intervening extragalactic magnetic fields.  Such fields might bend these
particles by as much as 5$^\circ$ or more from the original source direction.
Furthermore, it was tacitly assumed that the particle acceleration process
operates continuously to generate energetic particles. If instead, these
particles are created in short bursts, the cosmic ray arrivals will surely lag
the gamma-ray photons since they will be delayed by the additional path length
due to magnetic curvature so that no follow-up observation can succeed.  In the
former case with small magnetic deflections and constant flux, we can make 
some comparisons with theoretical estimates. Protheroe
and Stanev (Protheroe \& Stanev 1996) have estimated particle fluxes from the
decay of GUT-scale particles with masses $\simeq 10^{14}$ GeV.  Their results
for gamma-rays and protons are shown in figure 4 taken from their paper.  The
point at $3 \times 10^{11}$ GeV shows the cosmic ray flux inferred from the
single highest energy Fly's Eye event, averaged over 4$\pi$ steradians.  We do
not know if this one event comes from a particularly bright compact source or
simply represents one count from an otherwise isotropic distribution.  However,
a second event with an energy of $1.2 \times 10^{11}$ GeV measured by the
Yakutsk Array {\em et al.} (Efimov {\em et al.} 1991) was detected at
coordinates less than 8$^\circ$ away. Similar correlations have been observed
by Hayashida {\em et al.} (Hayashida {\em et al.} 1996).  Thus, the ultra-high
energy cosmic ray sky may in fact be highly anisotropic, a major focus of
interest for the proposed AUGER experiment (Auger Collaboration 1996).  If
true, the Fly's Eye flux could be directly compared with the TeV gamma-ray flux
limits determined above.  In figure 4, the Fly's Eye event at E = $3 \times 
10^{20}$
eV has been converted to a flux value by assuming a $4\pi$ steradian solid
angle and an energy bin $7.3 \times 10^{19}$ eV wide.  To compare with the Whipple gamma-ray
flux limits obtained here, we can similarly divide by $4\pi$ steradians and an
effective energy bin of 350 GeV (from the assumption of a constant $\nu 
F(\nu)$ spectrum). This is depicted as the lower of the two Whipple limits in
figure 4. An alternative upper limit can be derived from the general upper 
limit derived from the Whipple experiment on diffuse gamma rays.  Based on 
Monte Carlo simulations of the off-axis efficiency, the effective solid 
angle-collection area of the detector for diffuse gamma rays (or electrons) 
is 48 m$^2$-sr (Connaughton {\em et al.} 1997b).  The number of events 
passing the gamma ray selection criteria is $< 0.2$ per minute leading to an 
upper limit of 150 (E/1 GeV)$^{-2.7}~ photons/m^2-s-str$.  This is plotted 
in figure 4 as the greater of the two Whipple limits.  The results for 
Whipple and
HEGRA (Karle 1995) are displayed at energies of 350 GeV and 100 TeV
respectively.  Although the Whipple limit is about four times larger than the 
HEGRA result, it is 20 times closer to the gamma-ray estimates of Protheroe 
and Stanev.  One might be concerned that the TeV photons migrating to Earth
may be considerably scattered from the original ultra-high energy particle
direction. However the threshold energy is sufficiently low that relatively few
gamma-rays will be produced with transverse momenta of 1 GeV or more with
respect to the primary direction and thus outside the acceptance of our
gamma-ray selection criteria. 

\section{Emission from 3C147}
The most conservative assumption of the origin of the highest
energy cosmic rays is that they are accelerated in the jets of
Active Galactic Nuclei (AGN). The detection of TeV gamma-ray
emission from the two AGNs (Markarian 421 and 501) (Punch et al.
1992; Quinn et al. 1996) supports the notion of high energy
particle acceleration within some AGN. However within the error box
of the Fly's Eye event there is no AGN within 50 Mpc.

The most interesting extragalactic object within the error box is
the AGN, 3C147, one of the earliest optical quasars
discovered ({\em z} = 0.545). It is also very bright in radio and X-rays 
(luminosity in both bands in excess of $8\times10^{44}~ergs-s^{-1}$). It also
has a strong Faraday rotation. Thus, apart from its redshift, this
object is a prime candidate for identification as the source of the
Fly's Eye event and possibly the nearby Yakutsk event. 

If 3C147 was the source of the high energy particles, then it is
likely that it would also be a source of TeV gamma-rays. However
the redshift of 3C147 would suggest that there might be
considerable absorption of TeV gamma-rays by pair production on
infrared photons in intergalactic space. (Nikishov 1962; Gould 1966; Stecker 
1992, Biller 1995)  Observations of 3C147
(and two other AGNs) were made in the 1963-64 observing season in
Glencullen, Ireland by a combined Irish-U.K. team using a small
atmospheric \v{C}erenkov system with an energy threshold of 5
TeV (Long {\em et al.} 1964). A signal was detected at the 3 $\sigma$
level (corresponding to a flux of $1\times 10^{-10}$ photons-
cm$^{2}-s^1$). This would have indicated an incredible gamma-ray
luminosity of $5\times10^{47}$ ergs-s$^{-1}$. However this emission
was not verified and was not seen in any MeV-GeV gamma-ray
telescope experiment either.

3C147 (R.A.=05 39, Dec.=+49 49) was included in the survey with the
Whipple telescope reported above (G, Figure 2). In addition a
series of tracking observations were made with 3C147 in the center
of the field of view for maximum sensitivity. Six hours of
observation under optimum conditions gave no indication of a signal
and an upper limit of $1.8\times 10^{-11}$ photons-cm$^{2}-s^1$ was
derived. Given the greater sensitivity of the Whipple telescope it
appears most likely that the Glencullen result was a statistical
fluctuation. Hence there is no evidence from TeV gamma-ray
observations to support the identification of 3C147 as the cosmic
ray source.

\section{Discussion}
The experiment described above was an exploratory effort to correlate TeV
gamma-rays with UHE cosmic rays.  No evidence has been found for a steady
emission source at a flux level that might be expected for such an object. From
generic physics considerations, the co-production of gamma-rays and UHE
cosmic rays seems almost inevitable so an extension of these experimental
efforts is highly warranted.  By increasing the extent of the search fields and
the observation time, one could probe more deeply while providing a greater
margin for the unknown magnetic deflection of the UHE primary on its trajectory
to the Earth. 

The more likely scenario is that the processes that generate UHE cosmic rays
are episodic.  This makes the correlated detection of VHE gamma-rays
considerably more difficult since the cosmic rays will lag the photons by
intervals of the order of 100 years (Waxman \& Coppi 1996).  New detectors such
as MILAGRO and GLAST with wide fields of view are particularly suited for
investigating the existence of such transient gamma-ray fluxes.  More
generally, the study of short astrophysical transients has been barely
explored, even at optical wavelengths.

\acknowledgments

We acknowledge the technical assistance of Teresa Lappin and Kevin Harris. This
research is supported by grants from the US Department of Energy and by PPARC
in the UK, and by Forbairt in Ireland. 
\bigskip

\noindent
{\Large \bf References}
\bigskip

\begin{description}
\item{} Aharonian, F.A. {\em et al.,} 1992, Phys. Rev. {\bf D46,} 4188.
\item{} Akerlof, C.W. {\em et al.,} 1991, Ap.J. {\bf 377,} L97.
\item{} Auger Collaboration 1996, {\em Pierre Auger Project Design
Report,} October 31, 1995, Fermilab.
\item{} Bell, A.R., 1978, MNRAS {\bf 182,} 147.
\item{} Bhattacharjee, P., 1990, Proceedings of the ICRR
International Symposium, {\em Astrophysical Aspects of the Most Energetic 
Cosmic Rays,} ed.~M. Nagano \& F. Takahara (Singapore:  World 
Scientific), Kofu, Japan, 382.
\item{} Bhattacharjee, P. {\em et al.,} 1992, Phys. Rev. Lett. {\bf 69,} 567.
\item{} Biller, S., 1995, Astroparticle Physics {\bf 3,} 385.
\item{} Bird, D.J. {\em et al.,} 1994, Ap.J. {\bf 424,} 491.
\item{} Bird, D.J. {\em et al.,} 1995, Ap.J. {\bf 441,} 144.
\item{} Blandford, R.D. \& Ostriker, J.P., 1978, Ap.J. {\bf 221,} L29.
\item{} Buckley, J. {\em et al.,} 1997, A\&A. (submitted).
\item{} Connaughton, V. {\em et al.,} 1997a, Ap.J., {\bf 479,} 859.
\item{} Connaughton, V. {\em et al.,} 1997b, Astroparticle Physics (submitted).
\item{} Efimov, N.N. {\em et al.} 1991, Proceedings of the ICRR International 
Symposium, {\em Astrophysical Aspects of the Most Energetic Cosmic Rays,} 
ed.~M. Nagano \& F. Takahara (Singapore:  World Scientific), Kofu, Japan, 20.
\item{} Gould, R.J. \& Schreder, G.P.E., 1966, Phys. Rev. Letters {\bf 16,} 
252.
\item{} Greisen, K., 1966, Phys. Rev. Letters {\bf 16,} 748.
\item{} Hayashida, N. {\em et al.,} 1996, Phys. Rev. Letters {\bf 77,} 1000.
\item{} Karle, A. {\em et al.,} 1995, Physics Letters {\bf B347,} 161.
\item{} Legage, P.O. \& Cesarsky, C.J., 1983, A\&A {\bf 118,} 223.
\item{} Long, C.D. et al., 1964, Proc. I.A.U. Symposium \#23, Liege,
Belgium (Aug. 1964), Annales d'Astrophysique {\bf 6,} 251.
\item{} Milgrom, M. \& Usov, V., 1995, Ap.J. {\bf 449,} L37.
\item{} Miralda-Escud\'e, J. \& Waxman, E., 1996, Ap.J. {\bf 462,}
L59.
\item{} Nikishov, A.I., 1962, Sov. Phys. JETP {\bf 14,} 393. 
\item{} Protheroe, R.J. \& Stanev, T. 1996, Phys.~Rev.~Letters {\bf 77,} 3708.
\item{} Punch, M. {\em et al.,} 1992, Nature {\bf 358,} 477
\item{} Quinn, J. {\em et al.,} 1996, Ap.J. Letters {\bf 456,} L83
\item{} Reynolds, P. T. {\em et al.,} 1993, Ap.J. Letters {\bf 404,} 206
\item{} Sigl, G. {\em et al.,} 1994, Astropart. Phys. {\bf 2,} 401.
\item{} Stecker, F.W., DeJager, O.C., Salamon, M.H., 1992, Ap.J. Letters 
{\bf 390,} L49.
\item{} Vietri, M., 1995, Ap.J. {\bf 453,} 883.
\item{} Vietri, M. 1996, MNRAS {\bf 278,} L1.
\item{} Waxman, E., 1995, Phys. Rev. Letters {\bf 75,} 386.
\item{} Waxman, E. \& Coppi, P., 1996, Ap.J. {\bf 464,} L75.
\item{} Zatsepin, G.T. and Kuz'min, V.A., 1966, JETP Letters {\bf
4,} 78.
\end{description}

\begin{table}
\begin{center}
\begin{tabular}{ccc}
\tableline
\tableline
Position & Flux & Flux \\
& ($\times 10^{-11}$ erg cm$^{-2}$ s$^{-1}$) & ($\times 10^{-11} \gamma $ 
cm$^{-2}$ s$^{-1}$) \\
A & $3.5 \pm 1.2$ & $7.3 \pm 2.5$ \\
B & $3.6 \pm 1.0$ & $7.5 \pm 2.1$ \\
C & $4.4 \pm 1.1$ & $9.2 \pm 2.3$ \\
D & $2.7 \pm 1.1$ & $5.6 \pm 2.3$ \\
E & $5.1 \pm 2.3$ & $10.6 \pm 4.8$ \\
F & $7.7 \pm 3.1$ & $16.0 \pm 6.5$ \\
G & $3.0 \pm 2.2$ & $6.3 \pm 4.6$ \\
H & $4.9 \pm 1.5$ & $10.2 \pm 3.1$ \\
I & $3.6 \pm 1.2$ & $7.5 \pm 2.5$ \\
J & $3.9 \pm 1.4$ & $8.1 \pm 2.9$ \\
K & $1.4 \pm 2.2$ & $2.9 \pm 4.6$ \\
L & $1.5 \pm 1.3$ & $3.1 \pm 2.7$ \\
\tableline
\end{tabular}
\end{center}
\caption{Upper limits for E $>$ 350 GeV}
\label{tab:results}
\end{table}

\newpage

\newpage
\begin{figure}
\vspace{-1.0cm}
\caption{Bin excesses for Positions D to I.  The contours start
at $1 \sigma$ and increment in $1 \sigma$ steps.  With around 1000 bins per 
contour map, one might expect contours representing 2 or even 3 
$\sigma_{ij}$ deviations between the ON and OFF data.  The lack of such 
deviations may be explained by the fact that the OFF bin occupancy 
distributions are artificially smooth, each bin being an average of the 
equivalent bin in 6 to 9 different observations.
\label{fig:burst3}}
\end{figure}

\newpage
\begin{figure}
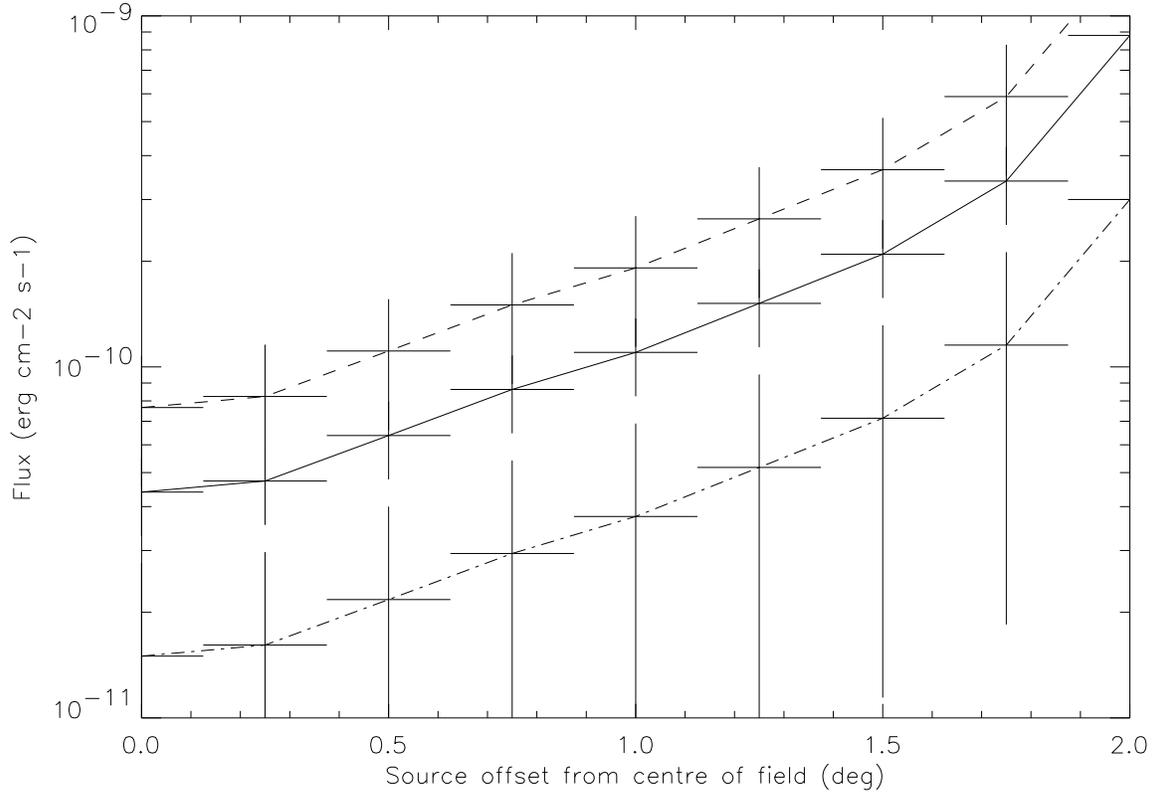

\caption{Upper limit to the flux above 350 GeV from topological defects for 
a point source over the field-of-view of the 10m reflector : Position C 
(solid), Position F (dashed) and Position L (dot-dashed).
\label{fig:tdeful}}
\end{figure}

\newpage
\begin{figure}
\caption{Plot of estimated $\gamma$-ray and proton fluxes from 
topological defects taken from Figure 1 of paper by Protheroe and
Stanev.  The Fly's Eye event at $3 \times 10^{20}$ eV is plotted with a 
filled circle.  The $\gamma$-ray upper limits are described in the text.  
The Whipple points refer to this experiment; the HEGRA data is from Karle 
(1995).
\label{fig:pro}}
\end{figure}

\newpage
\plotone{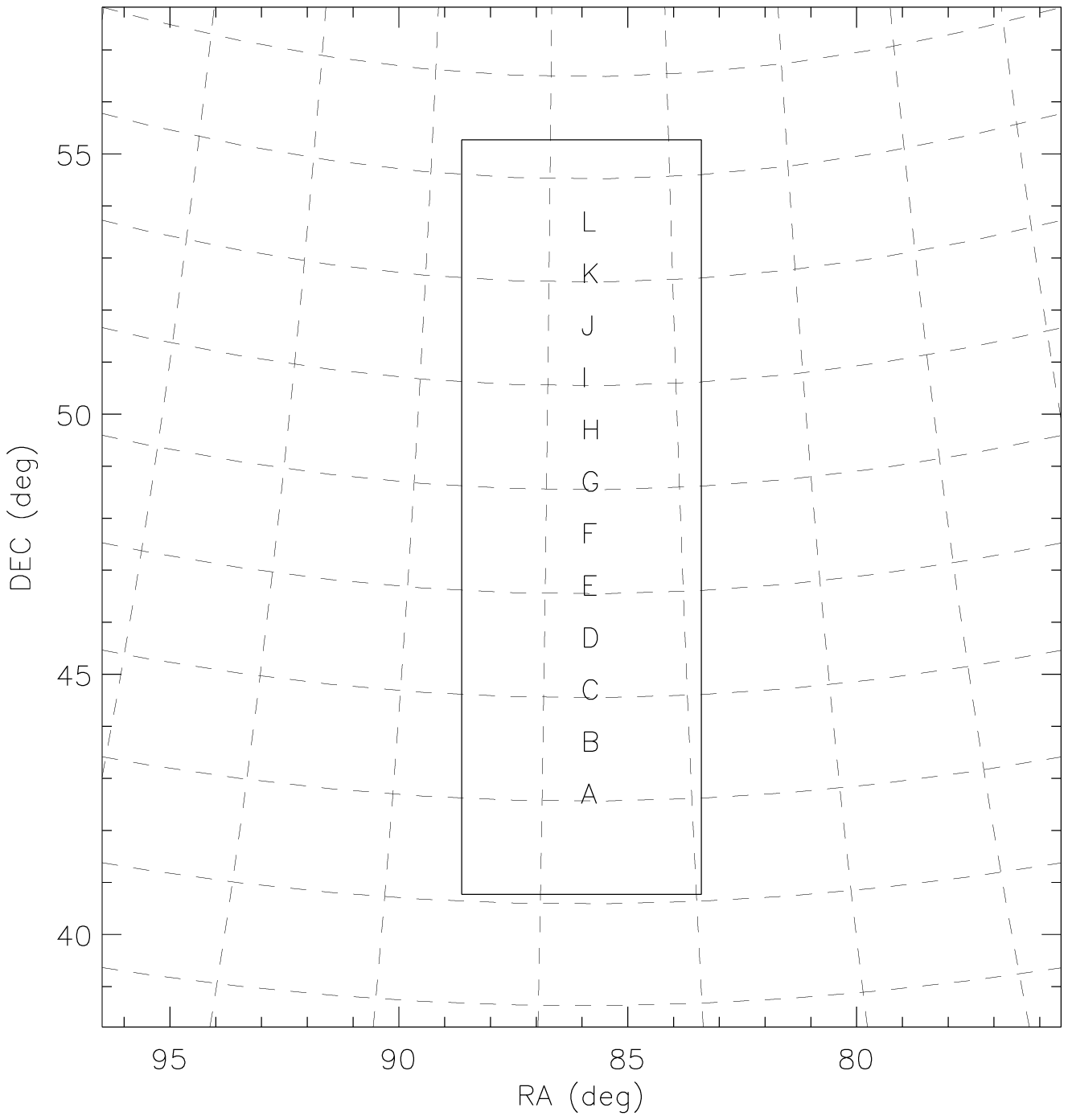}

\newpage
\vbox{
\plottwo{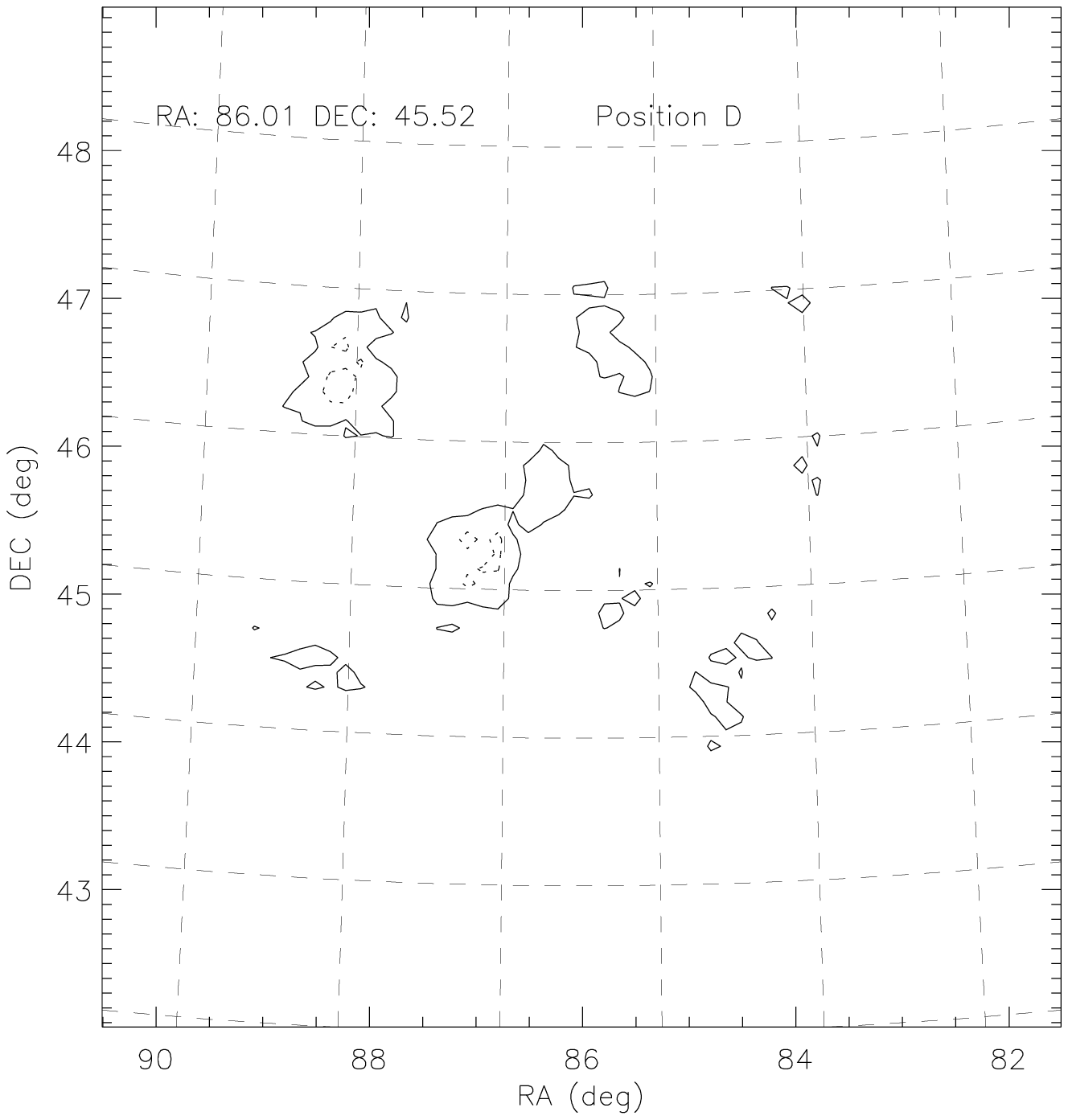}{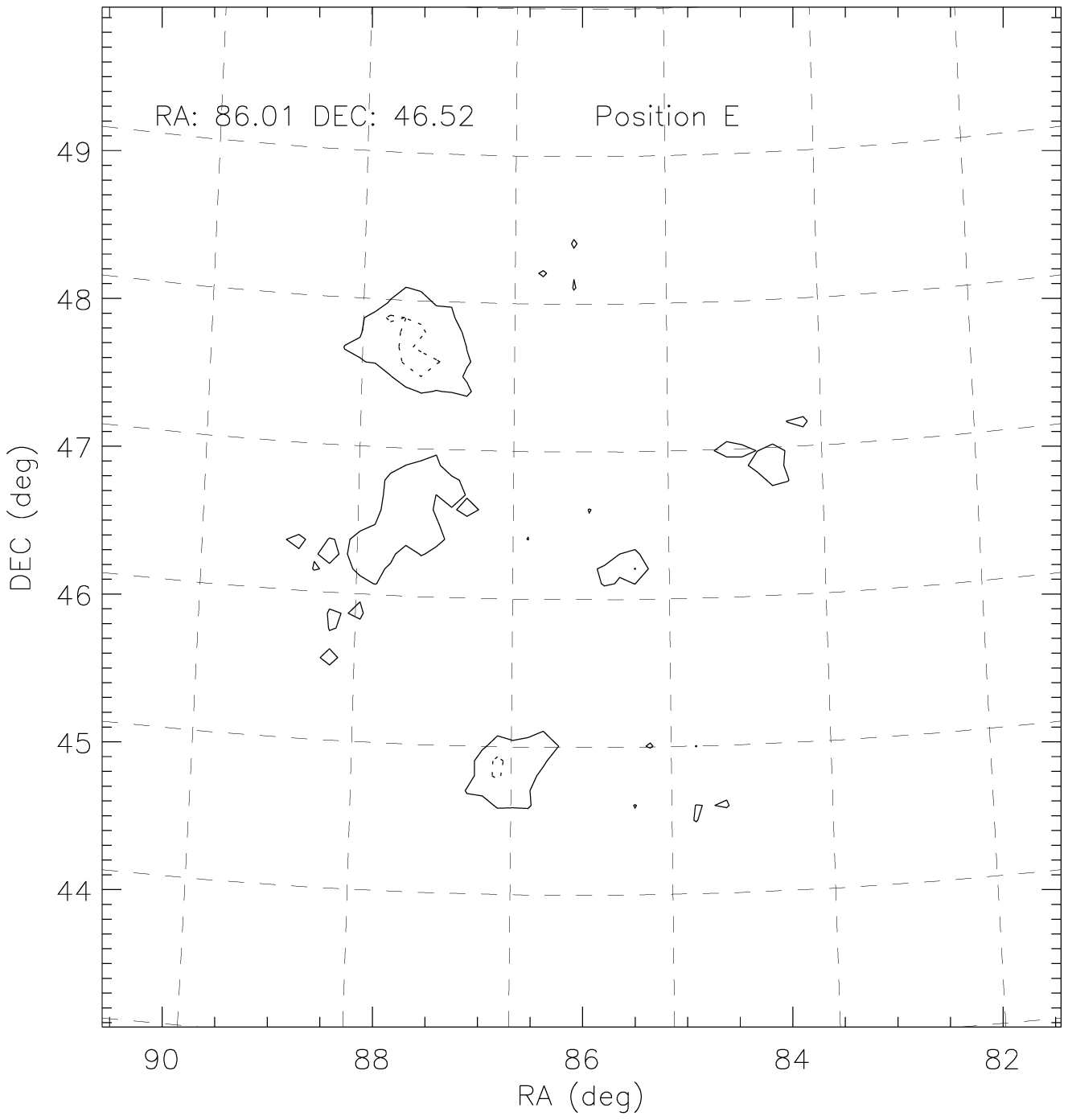}
\plottwo{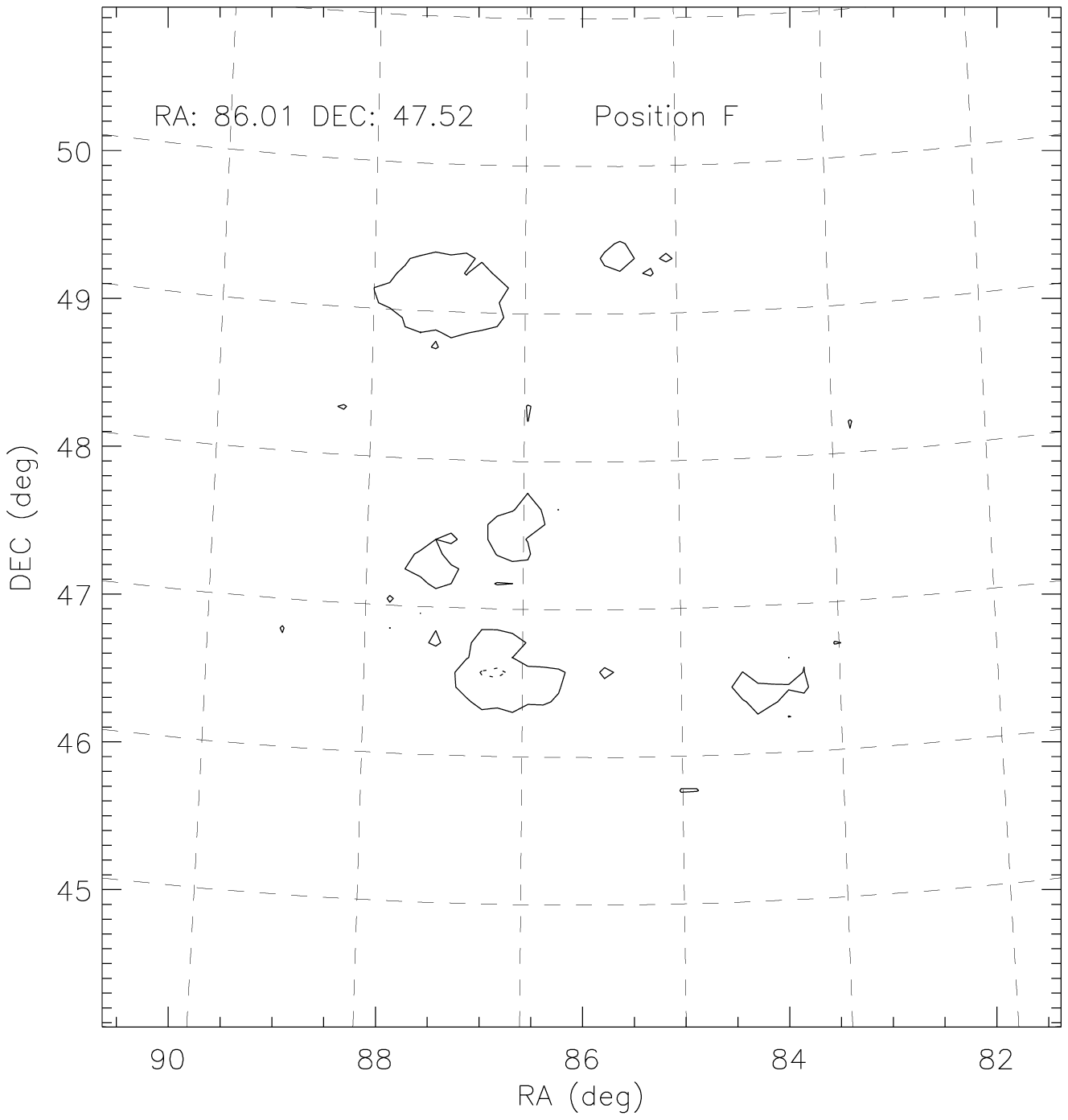}{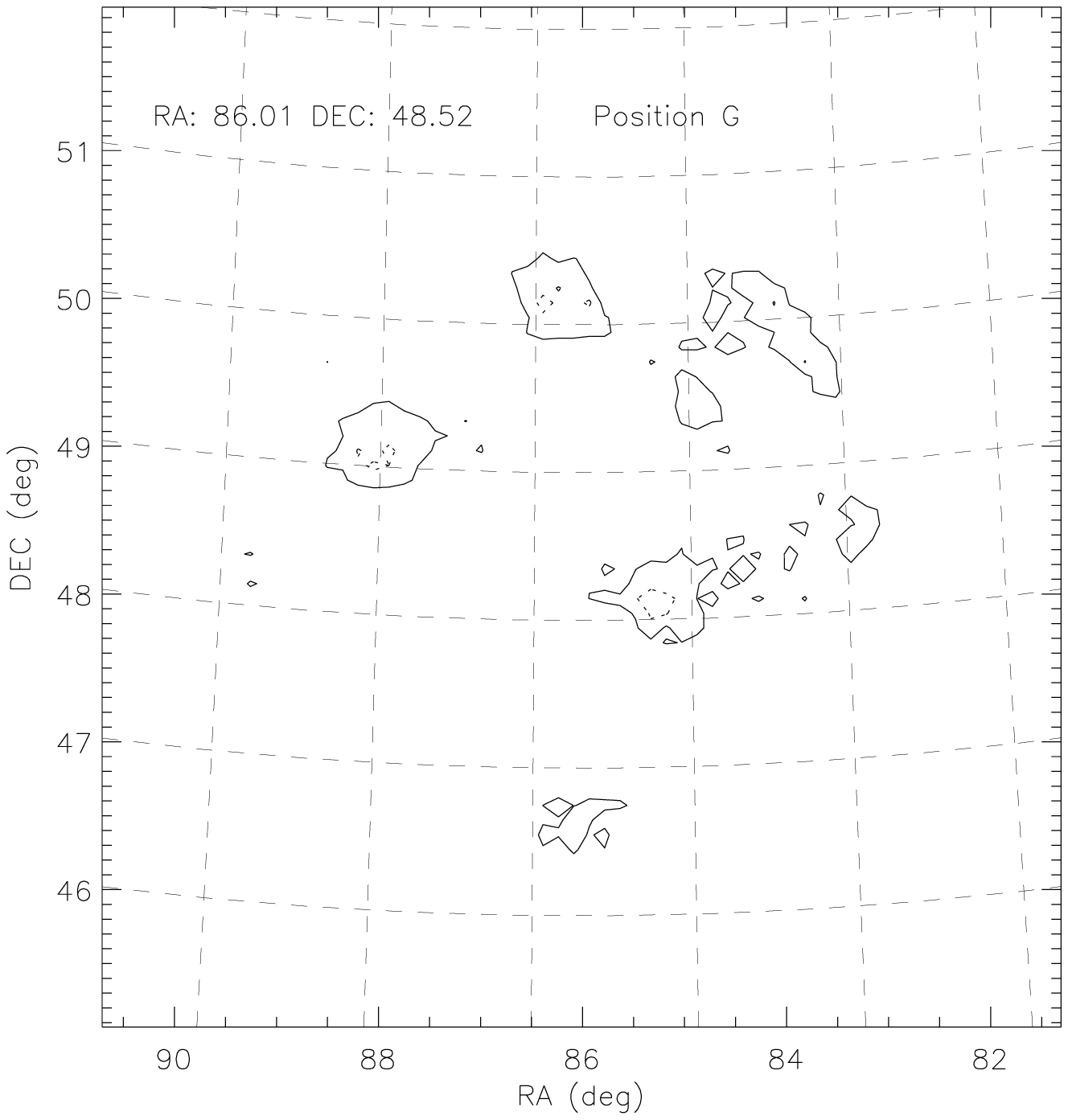}
\plottwo{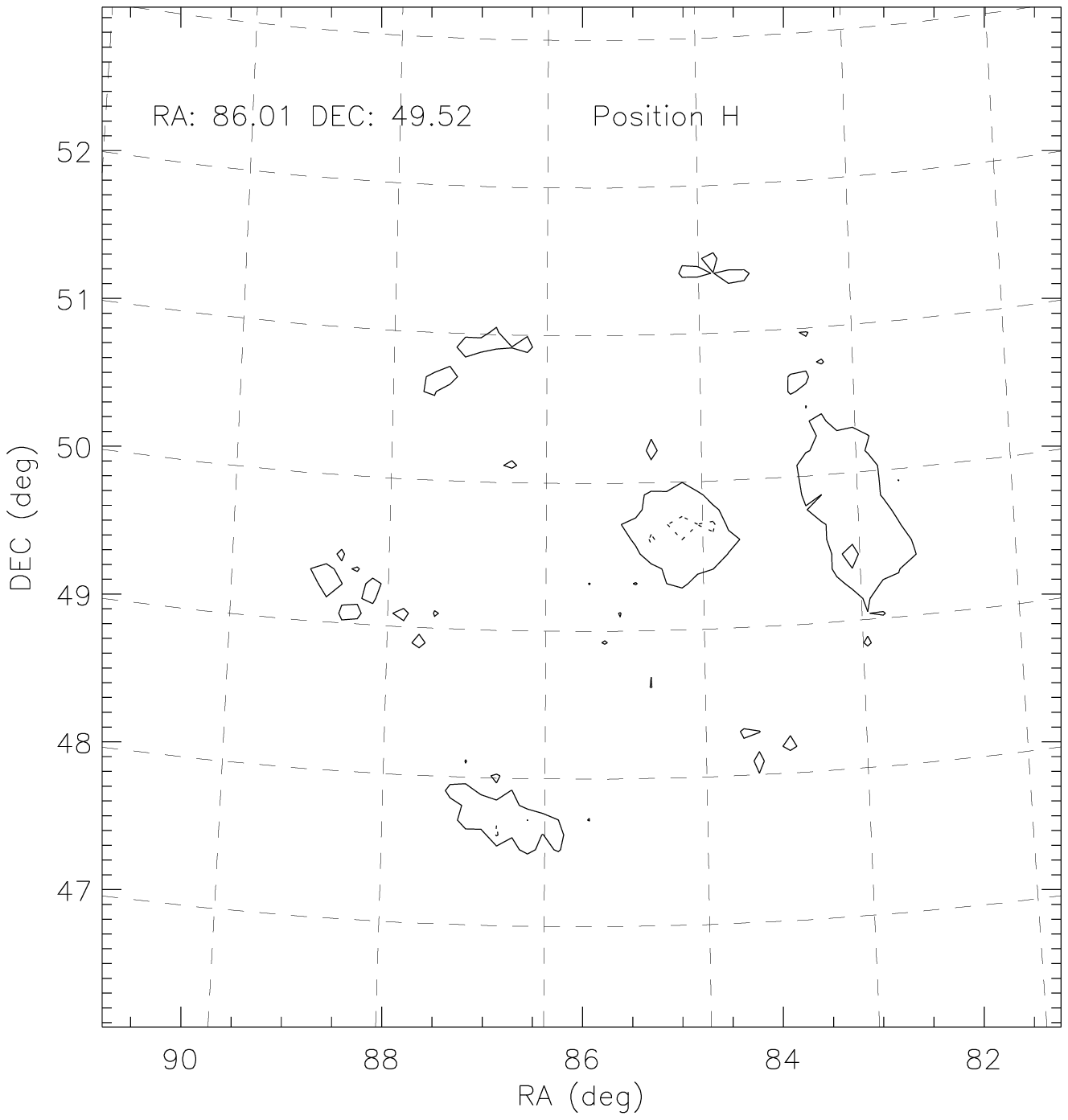}{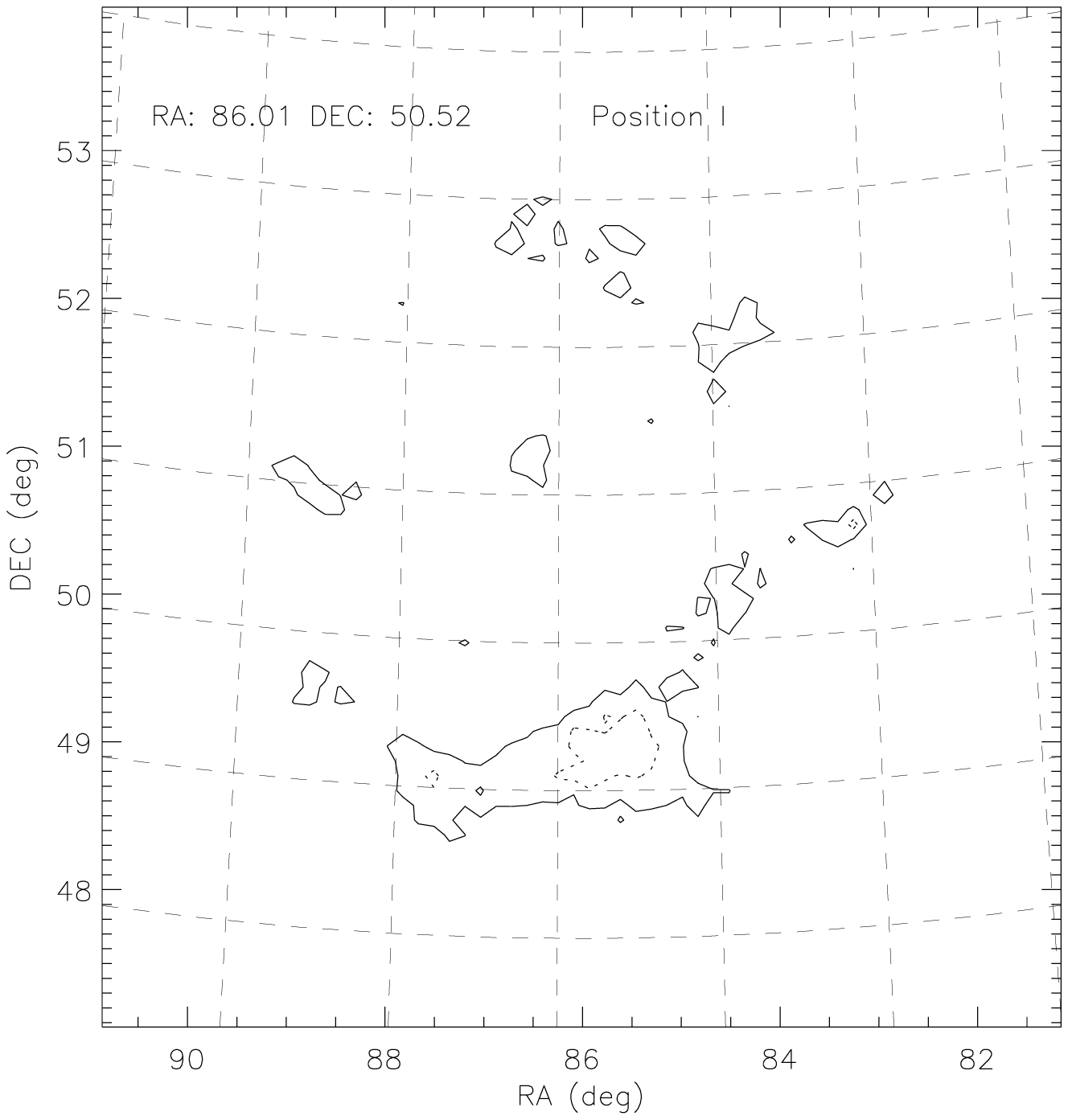}}

\newpage
\plotone{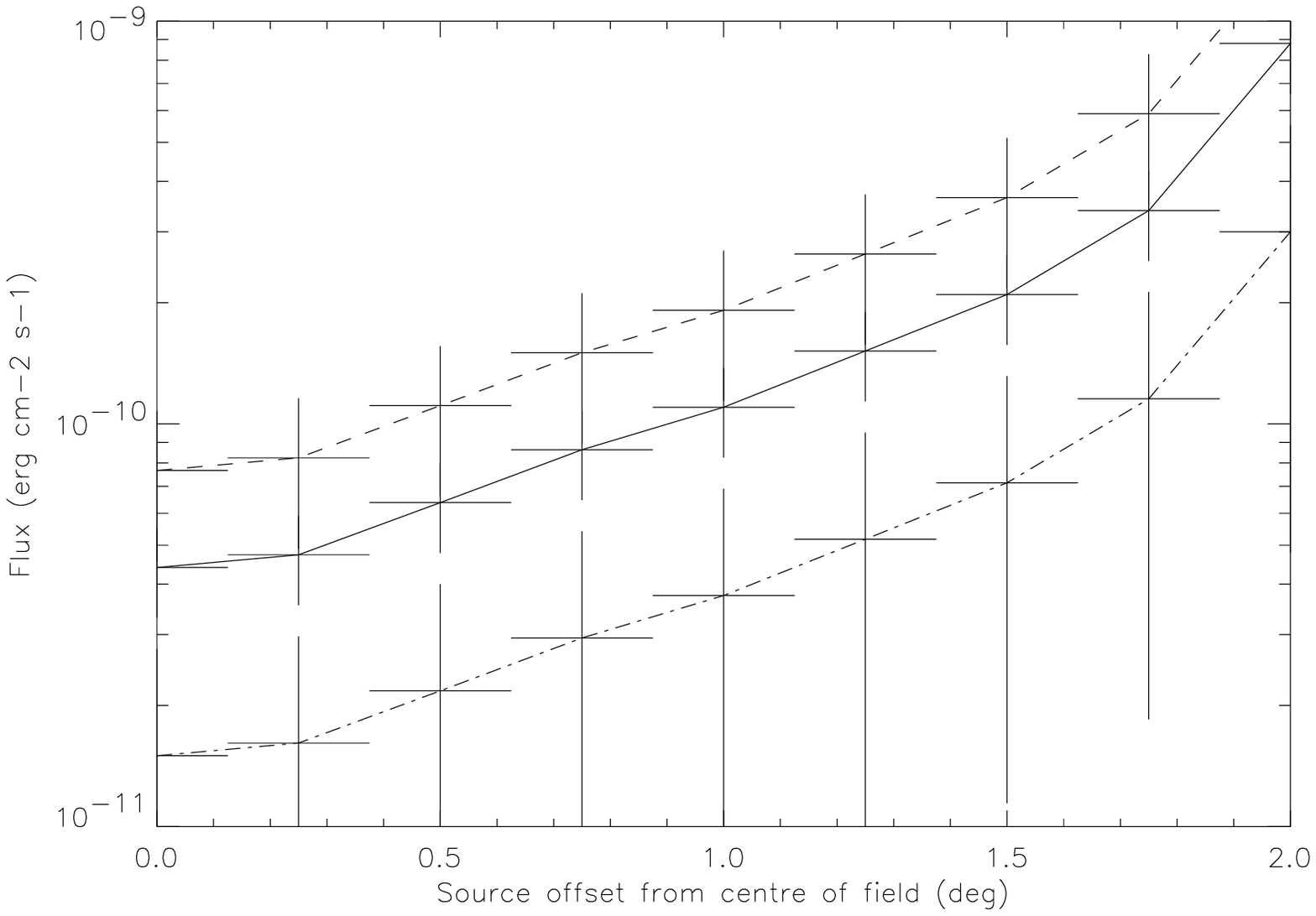}

\newpage
\plotone{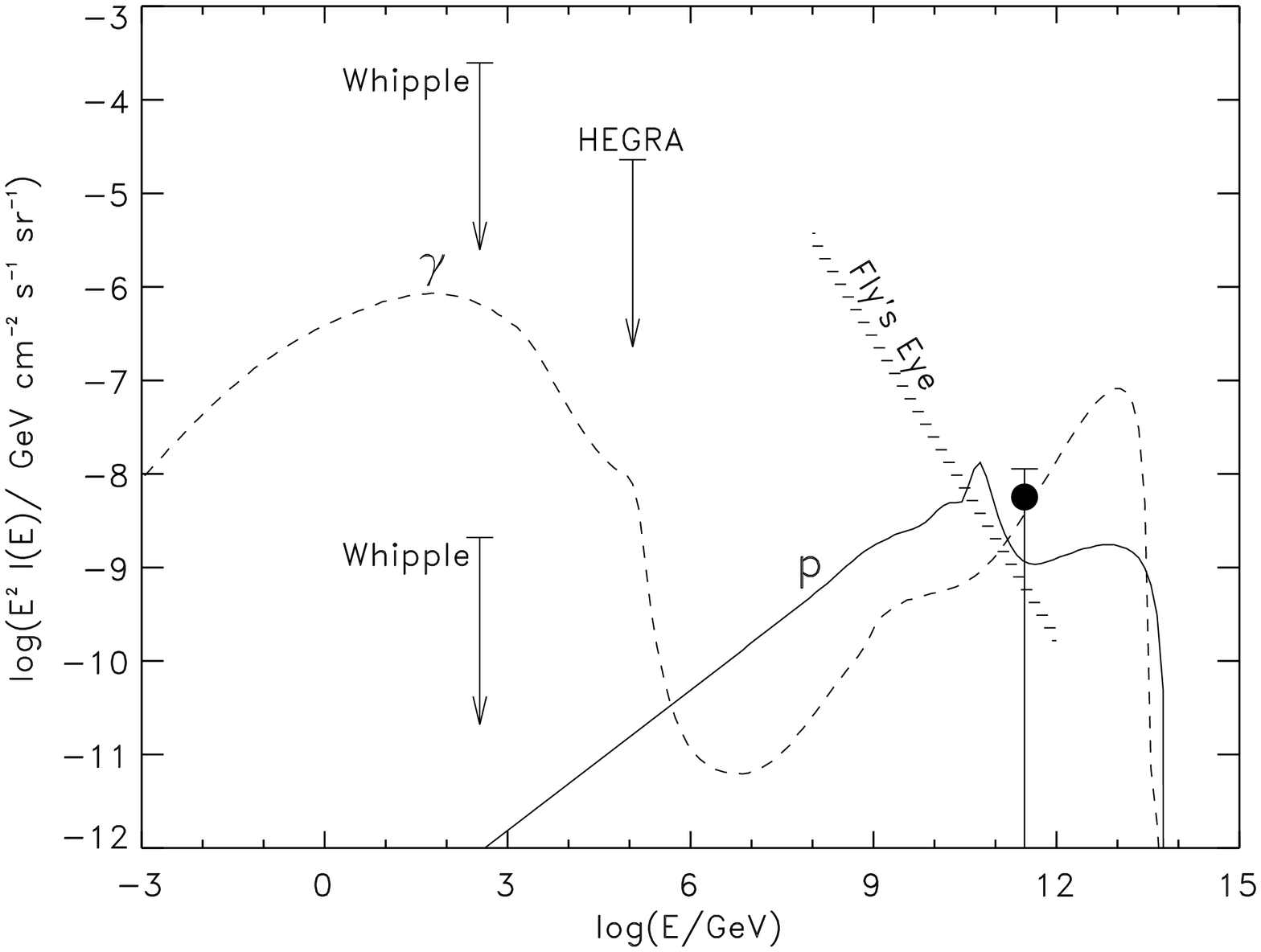}

\end{document}